\documentclass[10pt]{iopart}
\usepackage{epsf}

\begin{document}

\letter{Anomalous thermal conductivity and local temperature
distribution on harmonic Fibonacci chains} 

\author{Masashi Torikai and Takashi Odagaki
}

\address{Department of Physics, Kyushu University, Fukuoka 812-8581}


\begin{abstract}
The harmonic Fibonacci chain, which is one of a quasiperiodic chain 
constructed with a recursion relation, has a singular continuous 
frequency-spectrum and critical eigenstates. 
The validity of the Fourier law is examined for the harmonic Fibonacci 
chain with stochastic heat baths at both ends by investigating the 
system size $N$ dependence of the heat current $J$ and the local 
temperature distribution. 
It is shown that $J$ depends on $N$ as $J \sim (\ln N)^{-1}$ and the 
local temperature strongly oscillates along the chain. 
These results indicate that the Fourier law does not hold on the
harmonic Fibonacci chain. 
Furthermore the local temperature exhibits two different distribution 
according to the generation of the Fibonacci chain, i.e., the local 
temperature distribution does not have a definite form in the 
thermodynamic limit. 
The relations between $N$-dependence of $J$ and the frequency-spectrum, 
and between the local temperature and critical eigenstates are
discussed. 
\end{abstract}

\pacs{44.10.+i, 61.44.Br, 05.40.-a, 05.60.-k}



\nosections
Many studies in these decades have shown that arbitrarily defined
one-dimensional (1D) systems of interacting particles do not exhibit
normal thermal transport properties, i.e., the Fourier law does not hold
on such systems \cite{Rieder,Casher,Rubin,Dhar1,etc1}. 
For the steady-state of the homogeneous 1D chain of system size $N$, the 
Fourier law, $J=-\kappa \nabla T$ with thermal conductivity $\kappa$,
indicates that the heat current depends on the system size as $J \sim
1/N$ and that the temperature gradient $\nabla T$ is constant along the
chain.
Rieder \etal \cite{Rieder} have shown for a 1D chain of equal mass
particles interacting with identical harmonic potential that the heat
current is independent of the system size.
They also have obtained the local temperature distribution. 
The local temperature behaves in unphysical way:
the temperature takes a constant value in the bulk. 
Furthermore, near the end of the chain, the temperature decreases as we
move in the direction of the hotter heat bath, and rise only at the end
particle in contact with the heat bath;
the temperature exhibits corresponding behavior at the other end of the
chain. 
Casher and Lebowitz \cite{Casher} have shown for the same model but with
random mass distribution that $J \sim N^{-3/2}$.
For the same random mass distribution model but with different type of 
boundary conditions, Rubin and Greer \cite{Rubin} have obtained the
result $J \sim N^{-1/2}$.
These system size dependence of $J$ for periodic or disordered chains
may be attributed to the localization property of eigenstates on these
chains. 
Since the eigenstates are extended in periodic chains, the ballistic
energy transport of extended eigenstates results in the constant heat
current of the periodic chains.
In contrast, for the disordered chains, the decrease in heat current
with increasing system size is caused from the localized eigenstates,
which can not transport energy over the length of the system.
The extended and localized eigenstates correspond to continuous and
pure-point spectra, respectively.
It is shown that, for the Casher and Lebowitz type chain and heat bath,
the thermal conductivity $\kappa$ diverges as the system size increases
if the spectrum contains absolutely continuous part \cite{Casher}. 

Some of quasiperiodic systems have a Cantor set-like spectrum, i.e., a
singular continuous spectrum \cite{Kohmoto,Luck,Wurtz,etc2,etc3}. 
Then, on such systems, eigenstates show power law decay; such neither
extended nor exponentially localized states are called critical
states \cite{Kohmoto}. 
For example, a harmonic chain with quasiperiodically arranged spring
constants and/or mass of particles has a singular continuous spectrum
and critical eigenstates \cite{Luck,Wurtz,etc2}. 
We may thus expect that such quasiperiodic systems show exotic heat
transport properties compared with periodic or disordered systems.

In the present paper we investigate the anomaly of heat transport
phenomena on harmonic Fibonacci chains, which is one of discrete
quasiperiodic and self-similar 1D lattices.
We focus the anomaly resulting from the spectral properties of the
Fibonacci chain.
In order to check the validity of the Fourier law on the Fibonacci
chain, we investigate the system size dependence of the heat current
$J$.
Although Maci\'{a} \cite{Macia} have already studied the thermal
conductivity $\kappa$ of the harmonic Fibonacci chain, he has not
checked the system size dependence of $\kappa$ and thus the validity of
the Fourier law is not clear yet. 
Our results show that the heat current behaves as $J \sim (\ln N)^{-1}$,
which is in contrast with that of periodic or disordered chain. 
We discuss the fact that the total bandwidth of the phonon spectrum of
the Fibonacci chain has similar $N$ dependence to $J$. 
We also calculate the local temperature distribution on the Fibonacci
chain; it seems not to converge to a definite form even in the 
thermodynamic limit.
We relate $\{T_{i}\}$ with the critical eigenstates of the Fibonacci
chain. 

The harmonic Fibonacci chain which we consider is a 1D chain of $N$
particles; each particle interact with its neighbouring particles with
equal spring constant $k$.
We make the sequence of mass of particles $\{m_{i} |i=1,...,N ,
m_{i}=m_{\alpha} \mbox{ or }m_{\beta} \}$ according to the Fibonacci
sequence. 
The Fibonacci sequence of the $n$-th generation $L_{n}$, which consists
of two kinds of components $m_{\alpha}$ and $m_{\beta}$, is constructed
by the recursion relation: 
$L_{n}=L_{n-1}L_{n-2}$, $L_{0}=m_{\beta}$, and $L_{1}=m_{\alpha}$. 
Then the system size of the $n$-th generation Fibonacci sequence is the 
Fibonacci number $F_{n}$, which obeys the recursion relation
$F_{n}=F_{n-1}+F_{n-2}$, $F_{0}=1$, and $F_{1}=1$.
The Fibonacci number $F_{n}$ is asymptotically behaves as $F_{n} \sim
\tau^{n}$ with golden ratio $\tau=(\sqrt{5}+1)/2$.
We can obtain the asymptotic properties of the Fibonacci chain in the
limit of $N \to \infty$ by concerning the infinite-generation limit $n
\to \infty$. 
We set both the spring constant $k$ and the mass $m_{\beta}$ unity;
we calculate the heat current and the temperature distribution varying
with the mass $m_{\alpha}$. 

We consider the following Langevin equations for particles of the chain
with stochastic heat baths at both ends:
\begin{eqnarray}
 \eqalign{
 m_{1}\ddot{x}_{1}=-2x_{1}+x_{2}-\gamma \dot{x}_{1}+\eta_{\rm{L}}(t), \\ 
 m_{i}\ddot{x}_{i}=-2x_{i}+x_{i-1}+x_{i+1}, \\ 
 m_{N}\ddot{x}_{N}=-2x_{N}+x_{N-1}-\gamma \dot{x}_{N}+\eta_{\rm{R}}(t),}
\label{Langevin}
\end{eqnarray}
where $x_{i}$ are the displacements of the particles from their
equilibrium positions; $\gamma$ is the friction constant;
$\eta_{\rm{L}}$ and $\eta_{\rm{R}}$ are the random forces caused from
left and right heat baths, respectively.
We choose the random forces the white noises, i.e., the Fourier
transforms of the random force obey
\begin{eqnarray}
 \eqalign{
\langle \eta_{\rm{L}}(\omega) \eta_{\rm{L}}(\omega') \rangle
 =4 \pi \gamma T_{\rm{L}} \delta(\omega+\omega'), \\ 
\langle \eta_{\rm{R}}(\omega) \eta_{\rm{R}}(\omega') \rangle
 =4 \pi \gamma T_{\rm{R}} \delta(\omega+\omega'), }
\end{eqnarray}
where angular bracket is the average over the random force; $T_{\rm{L}}$
and $T_{\rm{R}}$ are the temperatures of the left and right heat baths,
respectively. 

The energy current $J_{i}$ from the site $i$ to the site $i+1$ is 
\begin{equation}
 J_{i}=\frac{1}{2} \langle (\dot{x}_{i+1}+\dot{x}_{i})(x_{i+1}-x_{i}) 
  \rangle.
\end{equation}
In the steady state, the energy current does not depend on the site;
then from the Langevin equation (\ref{Langevin}), 
\begin{equation}
 J=\Delta T \int_{0}^{ \infty}
  \frac{d\omega}{\pi} \frac{2 \gamma^{2}\omega^{2}}{|\det Y|^{2}},
  \label{heat-current} 
\end{equation}
where $\Delta T=T_{\rm{L}}-T_{\rm{R}}$ and $Y=\Phi-\omega^{2}M+\rmi \omega 
\Gamma$ with $\Phi_{ij}=-\delta_{i,j-1}+2\delta_{ij}-\delta_{i,j+1}$, 
$M_{ij}=m_{i}\delta_{ij}$, and
$\Gamma_{ij}=\gamma(\delta_{i,1}+\delta_{i,N})\delta_{i,j}$. 
We may write
\begin{equation}
 \det Y=D_{1,N}-\gamma^{2} \omega^{2} D_{2, N-1}
  +\rmi \gamma \omega (D_{2,N}+D_{1,N-1}),
  \label{detY}
\end{equation}
where $D_{i,j}$ denotes the determinant of the sub-matrix of
$\Phi-\omega^{2}M$ which begins with $i$-th row and column and ends with
$j$-th row and column \cite{Dhar1}. 

We define the local temperature at site $i$
\begin{equation}
 T_{i}=\left\langle p_{i}\frac{\partial H}{\partial p_{i}}
       \right\rangle 
 =\left\langle \frac{p_{i}{}^{2}}{m_{i}} \right\rangle,
\end{equation}
where $H$ is the Hamiltonian of the system.
We note that the equipartition relation $T_{i}=\langle x_{i} \partial H
/ \partial x_{i} \rangle$ holds for any mass distribution (for periodic
case, the equipartition relation has proven in \cite{Rieder}). 
For our model we can write the explicit form of the local temperature as
\begin{equation}
 T_{i}=T_{\rm{L}}-m_{i}\gamma \Delta T 
  \int_{-\infty}^{\infty}\frac{d\omega}{\pi} \cdot
  \omega^{2}
  \left|
  \frac{D_{1, i-1}+\rmi \gamma \omega D_{2, i-1}}{\det Y}
  \right|^{2}. \label{local-temperature}
\end{equation}

In calculating the heat current, it is convenient to express the
determinant $D_{i,j}$ in terms of a transfer matrix as 
\begin{equation}
 \left(
  \begin{array}{cc}
   D_{i,j}   & -D_{i+1,j}   \\
   D_{i,j-1} & -D_{i+1,j-1} 
  \end{array} 
 \right)
 =
 M_{j}M_{j-1} \cdots M_{i+1} M_{i},  \label{determinant}
\end{equation}
where the matrix $M_{i}(\omega)$ is the unimodular transfer matrix of
site $i$ defined as 
\begin{equation}
M_{i}(\omega)=
 \left(
 \begin{array}{cc}
  2-m_{i}\omega^{2} & -1 \\
  1                 &  0 
 \end{array}
 \right).
\end{equation}
If we rewrite the total transfer matrices of the $n$-th generation
Fibonacci chains as ${\mathcal M}_{n}=M_{N}M_{N-1}\cdots M_{1}$ with
$N=F_{n}$, then ${\mathcal M}_{n}$ obey a recursion relation 
\begin{equation}
\fl {\mathcal M}_{n}={\mathcal M}_{n-2}{\mathcal M}_{n-1},  \qquad
  {\mathcal M}_{0}=
 \left(
 \begin{array}{cc}
  2-m_{\beta}\omega^{2} & -1 \\
  1                 &  0 
 \end{array}
 \right)
 , \quad
  {\rm and}  \quad {\mathcal M}_{1}=M_{1}.
  \label{recursionM}
\end{equation}
Since we can obtain all the determinants $D_{i,j}$ needed in \eref{detY}
by calculating ${\mathcal M}_{n}$, the recursion relation
(\ref{recursionM}) saves a lot of time in calculating $J$. 

We calculated the heat current and the local temperature by numerical
integration of (\ref{heat-current}) and (\ref{local-temperature}),
respectively. 
As the system size grows larger, and as the mass ratio
$m_{\alpha}/m_{\beta}$ becomes farther from unity, the integrands of 
\eref{heat-current} and (\ref{local-temperature}) vary more rapid; 
hence we should use finer intervals on numerical integration.
For all the following data of $J$ and $\{T_{i}\}$ we checked that
decreasing the intervals of integration by a factor 5 did not make 
significant changes of the results. 

\begin{figure}
 \begin{center}
  \epsfbox{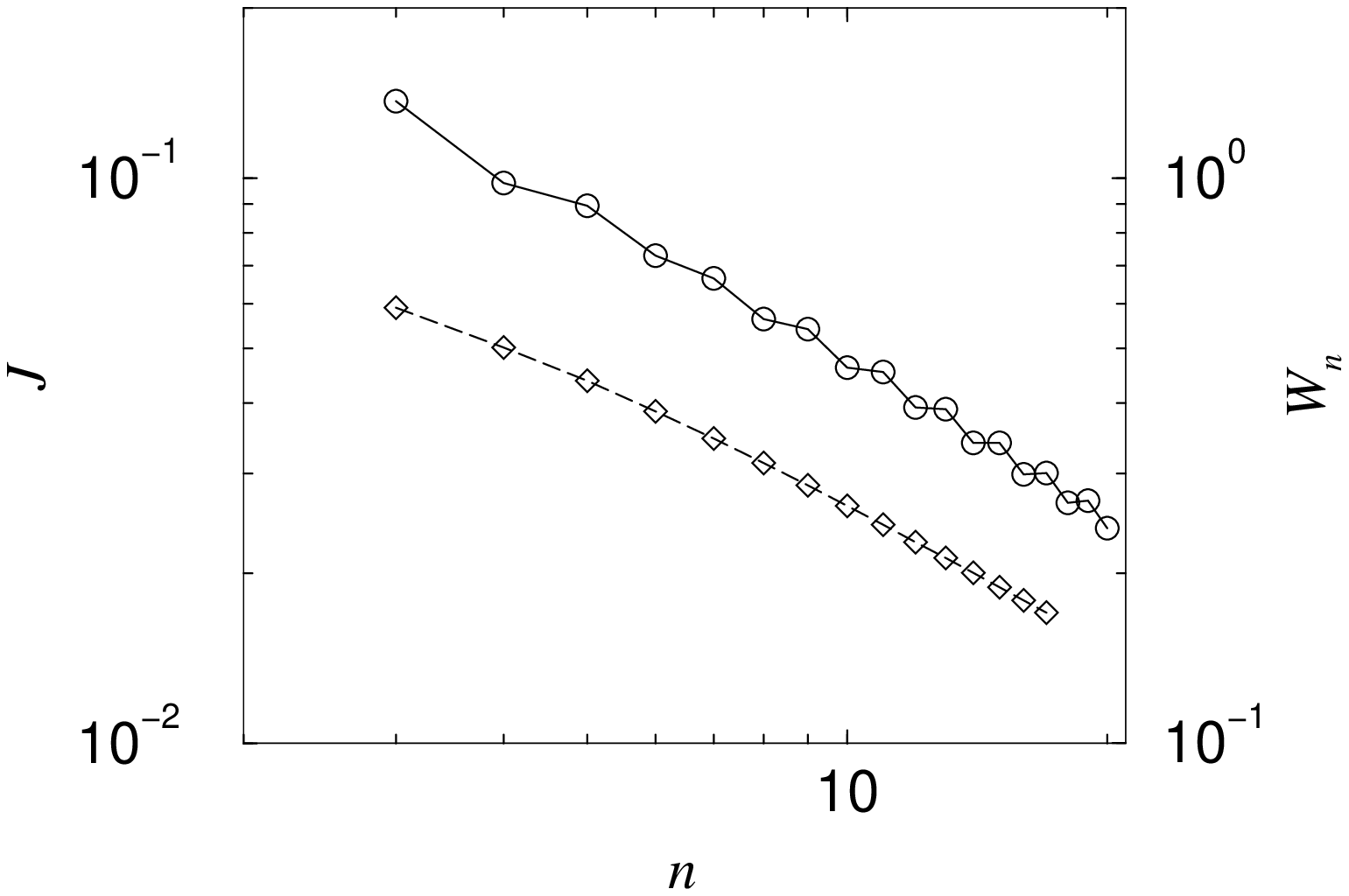}
 \end{center}
 \caption{ \label{fig1}
 Log-log plot of the generation $n$ versus the heat current $J$ (circle)
 and the total bandwidth $W_{n}$ (diamond). Lines are included to guide
 the eye. 
 The masses of particles in the Fibonacci chain are $m_{\alpha}=1.9$ and
 $m_{\beta}=1.0$. 
 The total bandwidth $W_{n}$ is normalized by the maximum
 eigenfrequency. 
 } 
\end{figure}

From the numerical calculation, we found that the total energy current 
$J$ decreases with oscillation as the generation $n$ increases (see
\fref{fig1}).
The decrease of $J$ for the generation $n$ shows power law behaviour 
$J \sim n^{-a}$; the exponent is $a \sim 1$ for sufficiently large
$m_{\alpha}$ (e.g., $a=1.03$ when $m_{\alpha}=3.0$ ).
This result indicates that it is appropriate to discuss the heat current
in terms of the generation $n$ rather than the system size $N$. 
We note that the system size dependence $J \sim n^{-a} \sim (\ln
N)^{-a}$ is the remarkable difference from the behaviour $J \sim N^{-b}$
of periodic or disordered systems.

\begin{figure}
 \begin{center}
  \epsfbox{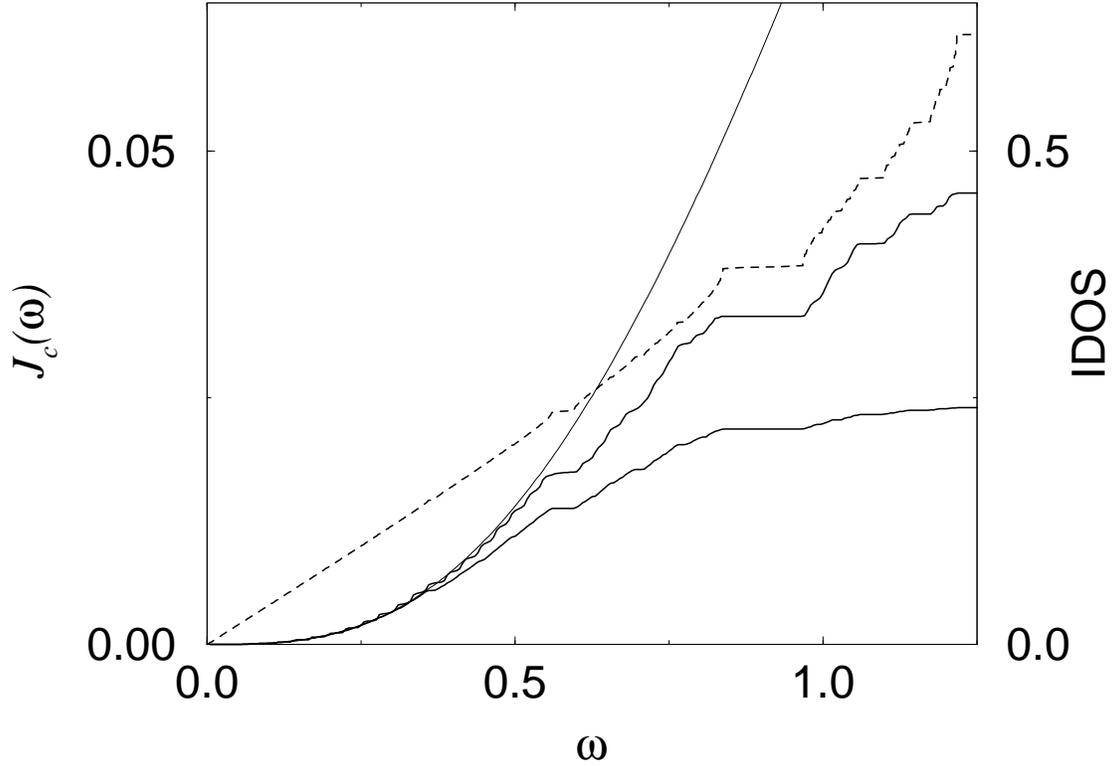}
 \end{center}
 \caption{\label{fig2}
 Full lines indicate the cumulative heat currents $J_{c}(\omega)$.
 The lower two full lines are $J_{c}(\omega)$ of 10th (upper line) and
 20th (lower line) generation Fibonacci chains with $m_{\alpha}=1.9$ and
 $m_{\beta}=1.0$. 
 The upper most smooth full line is $J_{c}(\omega)$ of a periodic chain
 with identical particles; the mass of particles is the average of
 particle in the above two Fibonacci chains. 
 Dashed line is the IDOS of the 15th generation Fibonacci chain.
 Although there are eigenstates in the high-frequency region (up to
 $\omega \sim 1.7$), we plot only low- and middle-frequency region
 because high frequency states contribute little to $J_{c}(\omega)$. 
 } 
\end{figure}

In order to discuss the contribution of eigenstates of the chain to the
total heat current we calculate, as done in \cite{Macia}, cumulative
heat current: 
\begin{equation}
J_{\rm c}(\omega)=\Delta T \int_{0}^{\omega}
  \frac{d\omega'}{\pi} \frac{2\gamma^{2}\omega'^{2}}{|\det Y|^{2}}. 
\end{equation}
The cumulative heat current $J_{c}(\omega)$ is a monotonically
increasing function;
the increment of $J_{\rm c}(\omega)$ in a certain $\omega$-region is the
contribution of the eigenstates of the $\omega$-region to the total heat 
current. 
For the Fibonacci chain, as shown in \fref{fig2}, $J_{\rm c}(\omega)$
consists of many slopes and plateaus.
Each slope consists of slopes and plateaus in nested fashion, and the
nested structure becomes finer as the generation grows; i.e., $J_{\rm
c}(\omega)$ seems to become the devil's staircase in the
infinite-generation limit.

The devil's staircase-like structure of $J_{\rm c}(\omega)$ is very
similar to the integrated density of states (IDOS) of the Fibonacci
chain (see \fref{fig2}). 
The plateaus of $J_{\rm c}(\omega)$ approximately coincide with the
plateaus of the IDOS; i.e., the frequencies corresponding to the
eigenstates make major contribution to the heat current. 

\Fref{fig2} shows that dominant contribution to $J$ comes from the low
frequency eigenstates, especially in high generation. 
We may attribute the high conductivity in the low-frequency region to
the extended eigenstates. 
The high frequency eigenstates are critical, i.e., they show a power-law
decay, while the low frequency eigenstates are extended like eigenstates
in periodic chains.
Correspondingly, the low frequency eigenstates may transport the energy
ballistically.
The ballistic transport can be seen from the similarity between
$J_{c}(\omega)$ of the Fibonacci chain and $J_{c}(\omega)$ of a periodic
chain in low frequency region (see \fref{fig2}). 
We should note that the low frequency region in which the eigenstates are
extended become narrower in the limit of infinite system size. 
This is because all the eigenstates become critical in the limit of 
infinite system size, since the continuous part of the spectrum of the
Fibonacci chain narrows and becomes singular continuous even in low
frequency region \cite{Luck}. 
As a consequence, we can conclude that the heat current, unlike that of
the periodic chain,  vanishes when $N \to \infty$. 

The similarity between $J_{c}(\omega)$ and IDOS indicates that the
generation dependence of $J$ is explained by the spectral properties of
the Fibonacci chain. 
Now, we consider the $n$-th generation rational approximants, which is a 
periodic chain whose unit cell is the $n$-th generation Fibonacci 
chain. 
We denote by $\sigma_{n}$ the $\omega^{2}$-spectrum of the $n$-th
generation rational approximant; 
$\sigma_{n}$ is given as $\{\omega^{2}; | \tr {\mathcal M}_n | \le
2. \}$. 
The spectrum $\sigma_{n}$ consists of $F_{n}$ bands;
as $n$ increases, the bands are fragmented into pieces and total
bandwidth $W_{n}$ decreases. 
We note that $W_{n}$ obeys a power law $W_{n} \sim n^{-a'}$ and $a' \sim
1$ \cite{Luck}. 
Since the frequency region corresponding to the bandgap does not
contribute to the heat current, we conclude that the decay of $J$ with
increasing $n$ is similar to the decay of $W_{n}$; thus $J$ obeys the
power law as $W_{n}$ does, as shown in \fref{fig1}.

Before showing the local temperature $\{T_{i}\}$, we define Fibonacci
sites and Fibonacci blocks, which seem important to discuss the behavior
of $\{T_{i}\}$. 
On the $n$-th generation Fibonacci chain, which consists of $F_{n}$
sites, the Fibonacci site is the $F_{j}$-th site with $j=1, 2, ..., n-1$ 
from left or right ends; i.e., the Fibonacci sites are numbered $F_{j}$
or $F_{n}-F_{j}$. 
We call a region between neighbouring Fibonacci sites the Fibonacci
block. 
Since $F_{j} \sim \tau^{j}$, Fibonacci sites are sparsely distributed
around the centre of the chain and densely near the ends. 
If the chain length is normalized to unity, the Fibonacci sites are
approximately at $\tau^{-j}$ and $1-\tau^{-j}$. 
Thus, as the generation increases, the Fibonacci sites distribution
becomes denser near the ends but invariant around the centre of the
chain. 
Correspondingly, the size of the Fibonacci blocks around the centre of
the chain stays constant in increasing generation. 

\begin{figure}
 \begin{center}
  \epsfbox{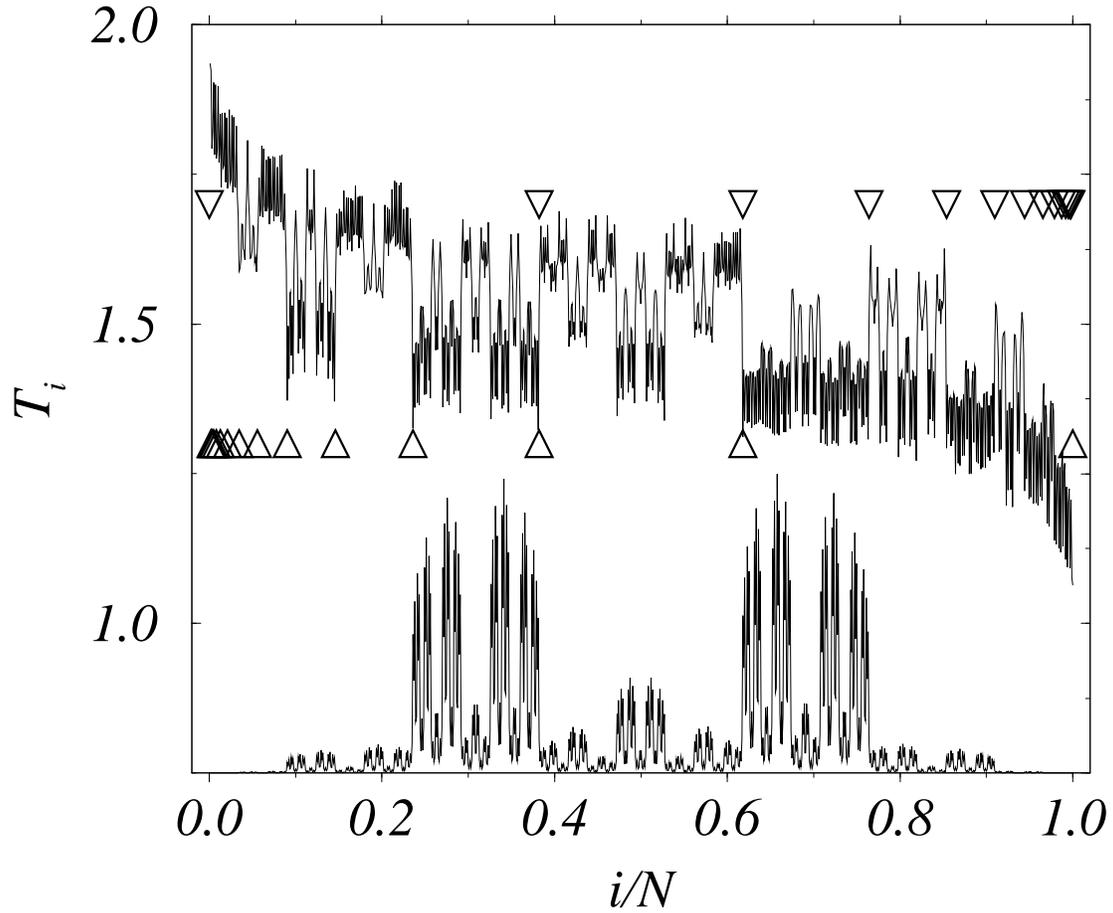}
 \end{center}
 \caption{\label{fig3}
 Upper full line is local temperature $\{T_{i}\}$ on the 15th generation
 Fibonacci chain with $m_{\alpha}=1.9$ and $m_{\beta}=1.0$.
 Triangles indicate the Fibonacci sites: upward and downward triangles
 correspond to $F_{j}$-th and $(F_{15}-F_{j})$-th sites, respectively. 
 Lower full line is $m_{i}x_{i}^{2}(\omega)$ of a critical eigenstate
 corresponding to the maximum eigenfrequency. 
 } 
\end{figure}

Now we plot $\{T_{i}\}$ with Fibonacci sites in \fref{fig3}.
We can see that the local temperature oscillates strongly along the
chain; especially $\{T_{i}\}$ has significant leaps at the Fibonacci
sites. 
Furthermore, the interior of each Fibonacci blocks have a bilaterally
symmetric temperature distribution.
The symmetry is a nontrivial result since the whole system with heat
baths at both ends is not bilaterally symmetric.
We note that some of the Fibonacci blocks have a symmetric mass
sequence, but not all the blocks do. 

\begin{figure}
 \begin{center}
  \epsfbox{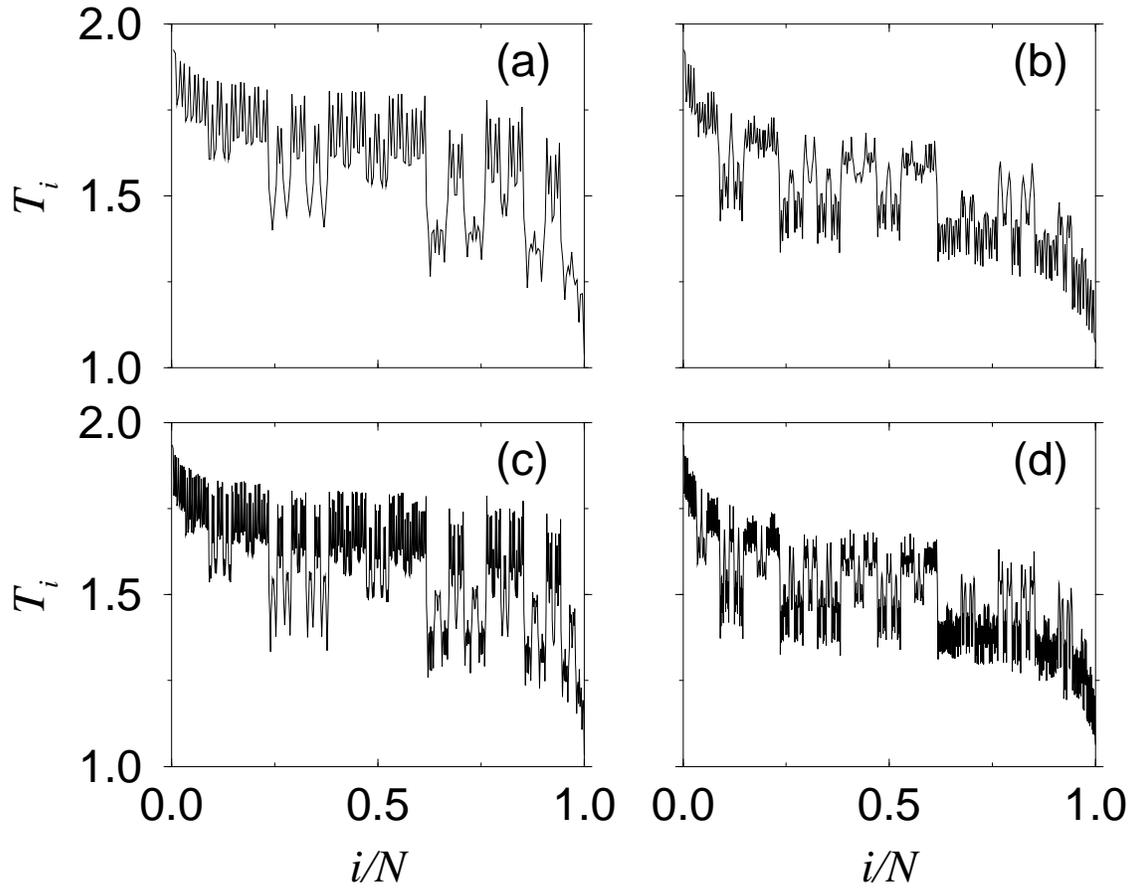}
  \end{center}
 \caption{\label{fig4}
 Local temperature on Fibonacci chain of (a) generation $n=12$ (number
 of particles $N=233$), (b) $n=13$ ($N=377$), (c) $n=14$ ($N=610$), and
 (d) $n=15$ ($N=987$).
 The Fibonacci chains consist of particles of $m_{\alpha}=1.9$ and
 $m_{\beta}=1.0$. 
 Each chain length is normalized to unity. 
 } 
\end{figure}

Another major characteristic of $\{T_{i}\}$ is an oscillatory behavior
for increasing generation.
We plot in \fref{fig4} the local temperature on the Fibonacci chains of
generation $n=12$, $13$, $14$ and $15$.
There is an obvious similarity between $\{T_{i}\}$ of $n=12$ and $n=14$,
and between $\{T_{i}\}$ of $n=13$ and $n=15$; i.e., $\{T_{i}\}$
oscillates periodically in generation with period two. 
We checked that the similarity in $\{T_{i}\}$ of odd- and
even-generation chains is maintained throughout the generations from
$n=6$ to $n=15$, which correspond to the system size from $N=13$ to
$N=987$. 
We may check the oscillation quantitatively by comparing the average
local temperature of, e.g., the centre Fibonacci block of each chain. 
The average temperature of the centre Fibonacci block is distributed
between 1.558 and 1.589 for odd-generation chains (from $n=7$ to
$n=15$); between 1.645 and 1.706 for even-generation chains (from $n=6$
to $n=14$). 
The periodic oscillation implies that the local temperature distribution
$\{T_{i}\}$ does not converge to a definite distribution even in the
limit of infinite-generation, i.e., in the thermodynamic limit.

It is interesting to note that the mass-weighted amplitude of critical
eigenstates also have leaps at the Fibonacci sites as shown in
\fref{fig3}. 
Such behaviour is similar to that of the $\{T_{i}\}$; thus we may
expect that the characteristic of the $\{T_{i}\}$ partly owes to the
critical eigenstates.
Since we do not observe a critical eigenstate with the symmetry in the
Fibonacci block, we may not attribute the cause of the symmetry of 
$\{T_{i}\}$ in Fibonacci sites to the properties of a single critical
eigenstate.

In summary, we have shown that the Fourier law is invalid for the
harmonic Fibonacci chain as for periodic or disordered harmonic chains. 
The heat current $J$ depends on the system size $N$ as $J \sim (\ln
N)^{-1}$;
the anomalous dependence is remarkably different from the periodic or
disordered case.
The $N$ dependence of $J$ is very similar to the $N$ dependence of the
total bandwidth of the frequency spectrum.
Such similarity is the consequence of the fact that only the lattice
vibration with frequency within the frequency-bands can transport the
heat over the length of the chain.
We have also obtained the local temperature.
The local temperature $\{T_{i}\}$ strongly oscillates and does not show
monotonic change.
Both $\{T_{i}\}$ and the mass-weighted amplitude of critical eigenstates
have significant leaps at Fibonacci sites;
this fact indicates the close relation between $\{T_{i}\}$ and
eigenstates.
We have concluded that $\{T_{i}\}$ does not converge to a definite form
in the limit of infinite system size since $\{T_{i}\}$ exhibits two
different form depending on the parity of the generation.


\Figures

\end{document}